\documentclass[preprint]{aastex}

\def \cm{~\rm{cm}}
\def \s{~\rm{s}}
\def \km{~\rm{km}}

\def \K{~\rm{K}}

\def \erg{~\rm{erg}}

\def \yr{~\rm{yr}}
\def \pc{~\rm{pc}}
\def \kpc{~\rm{kpc}}
\def \keV{~\rm{keV}}
\begin{document}

\title{CORRELATION OF AGN JET POWER WITH THE ENTROPY PROFILE IN COOLING FLOW CLUSTERS}
\author{Fabio Pizzolato\altaffilmark{1}, Tavish Kelly\altaffilmark{2},  and Noam Soker\altaffilmark{3}}

\altaffiltext{1}{INAF--Osservatorio Astronomico di Brera, Via Brera n.28,
        20121 Milano, Italy; fabio.pizzolato@brera.inaf.it}

\altaffiltext{2}{Department of Physics and Astronomy, Mississippi State University,
Mississippi State, MS 39762, USA; tck42@msstate.edu}

\altaffiltext{3}{Department of Physics, Technion -- Israel Institute of Technology, Haifa
32000, Israel;soker@physics.technion.ac.il}

\begin{abstract}
We find that the power of jets that inflate bubble pairs in cooling flow clusters of galaxies
correlates with the size $r_\alpha$ of the inner region where the entropy profile is flat,
as well as with the gas mass in that region and the entropy floor$-$the entropy
value at the center of the cluster.
These correlations strengthen the cold feedback mechanism that is thought to operate
in cooling flow clusters and during galaxy formation.
In the cold feedback mechanism the central super-massive black hole (SMBH) is
fed with cold clumps that originate in an extended region of the cooling flow volume,
in particular from the inner region that has a flat entropy profile.
Such a process ensures a tight feedback between radiative cooling and heating by the
SMBH (the AGN).
For a SMBH accretion efficiency (of converting mass to energy) of $\eta_j=0.1$, we find the
accretion rate to be
$\dot M_{\rm acc} \simeq 0.03 ({r_\alpha}/{10 \kpc})^{1.7}   M_\odot \yr^{-1}$.
This expression, as well as those for the gas mass and the entropy floor,
although being crude, should be used instead of the Bondi accretion rate when studying
AGN feedback.
We find that the mass of molecular gas also correlates with the entropy profile parameters
$r_\alpha$ and $M_g$, despite that the jet power does not correlate with the molecular gas mass.
This further suggests that the entropy profile is a fundamental parameter determining cooling and feedback
in cooling flow clusters.
\end{abstract}


\section{INTRODUCTION}
\label{s-intro}

It is now clear that in many clusters of galaxies moderate quantities of
the intra-cluster medium (ICM) are cooling to low temperatures ($T \ll 10^5 \K$;
see reviews by \citealp{Pete06} and \citealp{McNa07}).
{\it Moderate} implies here that the mass cooling rate to low temperatures
is much lower than the cooling rate expected without heating, but it is much larger than
the accretion rate onto the supermassive black hole (SMBH) at the center of the cluster.
The cooling mass is either forming stars (e.g., \citealp{Odea08}; \citealp{Raf08}),
forming cold clouds (e.g., \citealp{Edg10}), accreted by the SMBH,
or is expelled back to the ICM and heated when it is shocked.
In this \emph{moderate cooling flow model} (\citealp{Sok01, Sok03, Sok04}),
the SMBH is fed by cold clumps originating in an extended region ($r \sim 5 - 30 \kpc$)
of the cooling flow, in a process termed the \emph{cold feedback mechanism}
(\citealp{Piz05, Sok06, Piz07, Piz10}).
The cold feedback mechanism overcomes some severe problems encountered by feedback
models that are based on accreting gas directly from the hot phase (\citealp{Sok06, Sok09, Piz10}),
such as the problems in the Bondi accretion \citep{McNa10}.
The cold feedback mechanism is compatible with observations that show that
heating cannot completely offset cooling
(e.g., \citealp{Wis04, McN04, Cla04, Hic05, Bre06, Sal08, Wil09, Pete06}), and
preliminary steps are taken to include it in feedback simulations \citep{Gas10}.

The properties and behavior of the clumps that cool to low temperatures
were studied by us in previous papers \citep{Piz05, Sok06, Piz07, Piz10}.
The distribution of cold clumps is complicated, and there is no way for us to determine it.
For that, in this paper we use our recent results \citet{Piz10} that are summarized in
section \ref{s-basic}, to derive in section \ref{s-SMBH} a semi-empirical phenomenological
expressions for the accretion rate onto the SMBH.
The main result of \citet{Piz10} and the previous papers is the sensitivity of cold clump
evolution to the entropy profile.
Readers interested only in the phenomenological formulae for accretion can skip section \ref{s-basic}
and go directly to section \ref{s-SMBH}.
Our short summary is in section \ref{s-summary}.

\section{BASIC PROPERTIES OF COOLING CLUMPS}
\label{s-basic}

The cold feedback mechanism assumes that the ICM is populated by a widespread assembly
of cold clumps.
We define the clump's density contrast
\begin{equation}
\label{e-delta}
\delta = (\rho' -\rho)/\rho,
\end{equation}
where $\rho'$ is the mass density of the clump, and
$\rho$ is the mass density of the surrounding ICM.

As shown in \citet{Piz05, Piz10} the fate of a clump
critically depends on its initial overdensity.
$(i)$ {\emph{Stable clumps.} They start with a small initial overdensity, and
they are eventually stabilised.
$(ii)$ \emph{Relatively unstable clumps}. Moderately dense clumps, whose
density contrast decreases as they fall in, but they manage to reach the center.
$(iii)$ \emph{Absolutely unstable clumps}. These have $\delta > \delta_{C}$, and their
density increase monotonically.

The critical density is given by \citet{Piz10}
\begin{equation}
\label{e-chi}
\left[ \omega(\delta_{C}, T)\right] ^{-2}  \delta_{C} = \chi ,
\end{equation}
where
\begin{equation}
\label{e-omega}
\omega(\delta,T) =
\frac{(1+\delta)^{2}}{\Lambda(T)}  \Lambda\left(\frac{T}{1+\delta}\right) - 1,
\end{equation}
depends on $\delta$ and the ICM temperature $T$ through the cooling function $\Lambda$.
The left hand side of equation (\ref{e-chi}) is an implicit function of $\delta_{C}$, while the right
hand side depends on the ICM properties and the radius of the clump according to
\begin{equation}
\label{e-chi2}
\chi = \frac{3}{8} \; C_{D} \; \frac{g/a}{\omega_{\rm BV}^{4}\, \tau_{\rm cool}^{2}},
\end{equation}
where
\begin{equation}
\label{e-bv}
\omega_{\rm BV}^{2} = \frac{3}{5} \frac{g}{r} \frac{ d \ln K }{ d \ln r},
\end{equation}
is the Brunt--V\"ais\"al\"a frequency,
$\tau_{\rm cool}$ is  the ICM cooling time, $a$is the clump's radius (assumed spherical),
$g$ is  the gravitational acceleration, $K$ is the entropy,  and
$C_{D}\simeq 0.75$ is the drag coefficient \citep{Kai03}.
A plot of $\delta_{C}$ as a function of $\chi$ for several values of $T$
is shown in Fig.~\ref{f-deltac}.
Note that since $\chi \propto 1/a$, large clumps must be with a larger density
contrast than small clumps to be unstable.
\begin{figure}
\begin{center}
\vskip11mm
\includegraphics[scale=0.4]{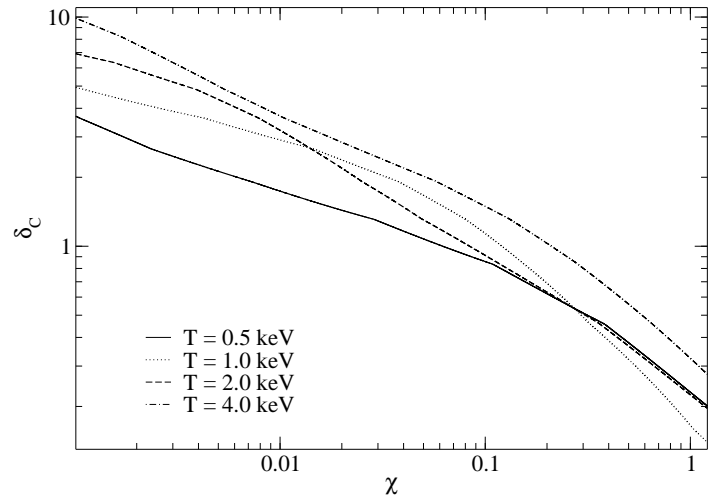}
\end{center}
\caption{\label{f-deltac}
Plot of the critical overdensity $\delta_{C}$, implicitly defined by
Equation  \ref{e-chi}  as a function of
the parameter $\chi$ (Equation~\ref{e-chi}) for four values of
the ambient temperature:
$T=0.5 \keV$ (solid line),
$T=1.0 \keV$ (dotted line),
$T=2.0 \keV$ (dashed line) and
$T=4.0 \keV$ (dot-dashed line).
}
\end{figure}

From equation (\ref{e-chi2}) it is clear that the evolution of overdense, hence cooler,
clumps is very sensitive to the cooling time and the entropy profile.
Cooling of clumps is favored for a short cooling time and a flat entropy profile
(by flat we refer to a moderate slope as well).
The sensitivity of the clump evolution to the entropy profile is
at heart of our present study.

We define the entropy logarithmic gradient
\begin{equation}
\label{e-alpha}
\alpha \equiv  \frac{ d \ln K }{ d \ln r},
\end{equation}
\noindent According to \citet{Don06} and \citet{Cav09} in cooling flow clusters
the entropy profile is flat, $\alpha \ll 1$, in the center, and it has $\alpha \simeq 1$ in the outer regions.
As in previous papers we take $10  \la a \la 100 \pc$.
For $\delta_C =1$ the condition for the clumps to be unstable is
$\chi \ga \chi_c \simeq 0.1$, where $\chi_c$ is the critical value of $\chi$.
Indeed, we define $r_\alpha$  to be the radius where $\alpha=0.5$,
and find that for our cluster sample $0.1 \la \chi(r_\alpha) \la 3$.
This condition on $\chi$ translates to a condition on the cooling time, evaluated at $\sim r_\alpha$
\begin{equation}
\label{e-tcool}
\tau_{\rm cool} \la 3 \times 10^8
\left( \frac {r}{10 \kpc} \right)
\left( \frac {a}{0.01r} \right)^{-1/2}
\left[ \frac {(gr)^{1/2}}{600 \km \s^{-1}} \right]^{-1/2}
\left( \frac{\alpha}{0.5} \right)^{-1}
{\chi_c}^{-1/2}
\yr.
\end{equation}

It is interesting to note that \citet{Raf08} found that in clusters where the cooling time
at a radius of $12 \kpc$ is $\tau_{\rm cool} (12 \kpc) < 8.5 \times 10^8 \yr \equiv \tau_s$,
high rate of star formation occurs.
Equation (\ref{e-tcool}) above gives a limit of $\sim 0.3 \tau_s$, but
for clumps of size $a=10^{-3} r$ ($a=10 \pc$ at $r=10 \kpc$) it gives a time scale of $\tau_{\rm cool} \simeq \tau_s$.
We conclude that a high cooling rate of over-dense clumps in the outer regions
of the cooling flow, where $\alpha \simeq 0.5-1$, can occur, according to the cold feedback mechanism,
when the cooling time is $\tau_{\rm cool} (10 \kpc) \la 10^9 \yr$, in agreement with observations
of star formation \citep{Raf08}.

In the inner regions where $\alpha \ll 1$, more clumps can cool. However, this region
contains less mass than the outer regions, and it is more prone to AGN heating.
Although there are much to be done, the cold feedback mechanism can generally account
for the limit of $\tau_s \simeq 10^9 \yr $ for a substantial star formation to occur,
as found by \citet{Raf08}.

It is observed that major AGN activity occurs at a general time intervals of
$\sim 10^8 \yr$, (e.g., \citealp{Wis07} for Hydra A).
In the cold feedback mechanism the time scale is determined mainly by regions
from where large quantity of gas might be cooling.
This occurs where the entropy profile becomes flat (moving inward),
$r_\alpha \simeq 3 - 30 \kpc$ (\citealp{Don06}; \citealp{Cav09}).
The free fall time from these regions is $\tau_{ff} \sim 10^7 - 10^8 \yr$.
The duty cycle might take somewhat longer.
We attribute the time intervals between major AGN eruptions to the time it takes
unstable clumps in large numbers to fall from these regions onto the SMBH.

\citet{Sok08}  considered both the duty cycle and the radiative cooling time
 $\tau_{\rm cool} (12 \kpc) < 8.5 \times 10^8 \yr \equiv \tau_s$
condition for high star formation rate in the frame of the cold feedback mechanism.
\citet{Sok08} proposed the criterion that the feedback cycle period must be
longer than the radiative cooling time of dense blobs for large quantities of gas
to cool to low temperature. It is possible that the two conditions are required for star formation:
that cold blobs in large number are unstable and cool to low temperature, and that the
AGN heating duty cycle is longer than the cooling time.

\section{SMBH MASS ACCRETION RATE}
\label{s-SMBH}
As usually done, we take the power in the jets to be
\begin{equation}
\label{e-jets}
\dot E_{\rm jets} = \eta_j \dot M_{\rm acc} c^2.
\end{equation}
We take the energy of the jets to be equal to the energy in bubble pairs of clusters.
The cavity power is calculated assuming $4pV$ of energy per cavity, and the buoyancy timescale $t_{\rm b}$
for its formation.
The energy and power of bubbles are taken from the list compiled by \citet{McNa10}, with
complementary material from \citet{Bir04} and \citet{Raf06}.
The power of the jets is listed in the fifth column of Table 1.
The clusters listed in Table 1 are those that we could find both entropy profile and jet
(bubble; cavity) power in the literature.
On the one hand this is the minimum energy, as some of the jets' energy goes to excite shocks and sound waves.
On the other hand, we think the bubble filling is mainly non-relativistic, such that their energy
should be $(5/2 PV)$. Over all, the usage of $\dot E_{\rm jets} =P_{\rm cav}=4pV/t_{\rm b}$
for the power of cavities is adequate for the present purpose.
However, for the reasons listed above we think that the uncertainties in the power of
jets are much larger than the uncertainties listed by \citet{McNa10}.
In the present Letter we do not include the uncertainties in the analysis, which is equal to
assuming an equal (large) uncertainties to all points.

The entropy profiles, as well as some other cluster properties are
taken from the catalog compiled by \citet{Cav09}. From the entropy profile we find the
radius $r_\alpha$, listed in the second column of Table 1, defined to be the radius
where $\alpha=0.5$ (equation \ref{e-alpha}).
Inward to $r_\alpha$ the entropy profile is flat.
 We also define the gas mass parameter
\begin{equation}
\label{e-mg}
M_{g} \equiv \frac {4 \pi}{3} \rho(r_\alpha) r_\alpha^3,
\end{equation}
listed in the third column of Table 1.
We use this parameter rather than the total gas mass inside $r_\alpha$ because in many cases the
large bubbles near the center distort the density profile, and we are after a simple relation.
We also list values of $\chi(r_\alpha)$ for $a=0.01r$ and $gr_\alpha = (600 \km \s^{-1})^2 $ in the
fourth column of Table 1.
%
%
\begin{table}
\caption{DATA ON CLUSTERS}
\begin{tabular}{l c c c c c c }
\hline
Cluster & $r_\alpha$ & $M_g(r_\alpha)$ & $\chi$ & $P_{\rm cav}$            & $M_{\rm mol}$  & $K_0$  \\
  &   (kpc) & $(10^8 M_\odot)$&  & $(10^{42} \erg \s^{-1})$ & $(10^8 M_\odot$) & $({\rm keV} \cm^2)$   \\
\hline
A478        & 6.3    &  30     & 1.32    & 100    & $25 ^{a}$   &  7.8 \\
A1795       & 15.1   & 190     & 1.11    & 160    & $55^{b}$    & 19 \\
Perseus     & 20.8   & 500     & 2.78    & 150    & $350 ^{c}$  & 19.4 \\
A2199       & 6.2    &  12     & 0.24    & 270    & $2.7 ^{d}$  & 13.3 \\
A2052       & 7.4    &   5.9   & 0.11    & 150    &             &  9.5 \\
A4059       & 3.0    &   1.7   & 0.19    &  96    &             &  7.1  \\
Cygnus A    & 15.3   & 150     & 0.52    &3900    & $15 ^{d}$   & 23.6 \\
A2597       & 12.2   & 110     & 1.86    & 67     & $45 ^{a}$   & 10.6 \\
Hydra A     & 11.6   &  77     & 0.83    &2000    & $11^{a}$    & 13.3   \\
Centaurus   & 1.2    &   0.27  & 0.89    &   7.4  &             &  2.2  \\
RBS797      & 25.2   &1180     & 2.77    &1200    &             & 20.9 \\
2A 0335+096 & 8.9    &  50     & 3.19    &  24   &             &  7.1 \\
A133        & 14.6   &  83     & 0.71    & 620    &             &  17.3 \\
A262        & 5.1    &   3.4   & 0.14    &   9.7  & $9.30 ^{d}$ &  10.6  \\
M87         & 1.8    &   0.55  & 0.20    &   6.0  & $0.08 ^{e}$ &  3.5  \\
HCG62       & 2.7    &   0.99  & 0.92    &   3.90 &             &  3.4 \\
MKW3S       & 18.5   & 130     & 0.35    & 410    & $5.4 ^{d}$  &23.9\\
\hline
\end{tabular}

{Notes:
The meanings of the quantities are as follows.
$r_\alpha$ is the radius inward to which the entropy profile is flat, i.e., $\alpha < 0.5$.
The entropy profiles and entropy floor $K_0$ are from \citet{Cav09};
The gas mass parameter $M_g(r_\alpha)$ is defined in equation (\ref{e-mg}); $\chi$ is according to equation (\ref{e-chi2});
The power of the cavity pair, $P_{\rm cav}$, is calculated by
dividing the cavity pair energy, $4PV$, by the buoyant time, and taken from \citet{McNa10} and \citet{Raf06}.
The molecular mass $M_{\rm mol}$ sources:
(a) \citet{edge01}; (b) \citealp{sal04}; (c) \citet{sal06}; (d) \citet{sal03}; (e) \citet{tan08}.
}
\end{table}

We could not find a simple correlation between the jet power and $\chi$, nor with
the cooling time $\tau_{\rm cool}$, and nor with $M_g/\tau_{\rm cool}$.
A short cooling time of $\tau_{\rm cool} \la 10^9 \yr$ at $r \simeq 10 \kpc$
is a condition for high star formation rate \citep{Raf08}.
However, once this condition is met it is not clear if the cooling time directly
determines the mass accretion rate onto the SMBH.

In the cold feedback mechanism, the entropy profile seems to be a more fundamental
parameter for mass accretion rate onto the SMBH, once a short cooling time is established.
It was already found that high star formation rate and high $H\alpha$ luminosity
require that the entropy floor in the cluster be $K_0 = kTn^{-2/3} \la 30 \keV cm^{2}$ \citep{Voit08}.
We look for a correlation rather than a limit.
In Fig. \ref{fig:correlations} we plot the jets (cavity pair) power versus the size of the flat entropy
region $r_\alpha$ (upper panel), and the gas mass parameter $M_g$ (middle panel).
A correlation, albeit not very strong, is clearly seen in all panels of Fig. \ref{fig:correlations}.
We look for the simplest correlation in the logarithmic plot, namely, a linear one.
The correlations we find are
\begin{equation}
\label{e-corr}
\log P_{\rm cav}({\rm erg} \s^{-1}) \simeq 1.65 \log r_\alpha({\rm kpc}) + 42.6,
\end{equation}
with Pearson R coefficient of $R^2=0.67$,
 and
\begin{equation}
\label{e-cormg}
\log P_{\rm cav}({\rm erg} \s^{-1}) \simeq 0.58 \log M_g (M_\odot) + 38.5,
\end{equation}
with $R^2 =0.56$.
Although these are not tight correlations, when the large scatter that is expected from the
temporarily variations in the AGN power and its influence on the ICM is considered,
these correlations have a merit.
\begin{figure}
\begin{tabular}{c}
{\includegraphics[scale=0.28]{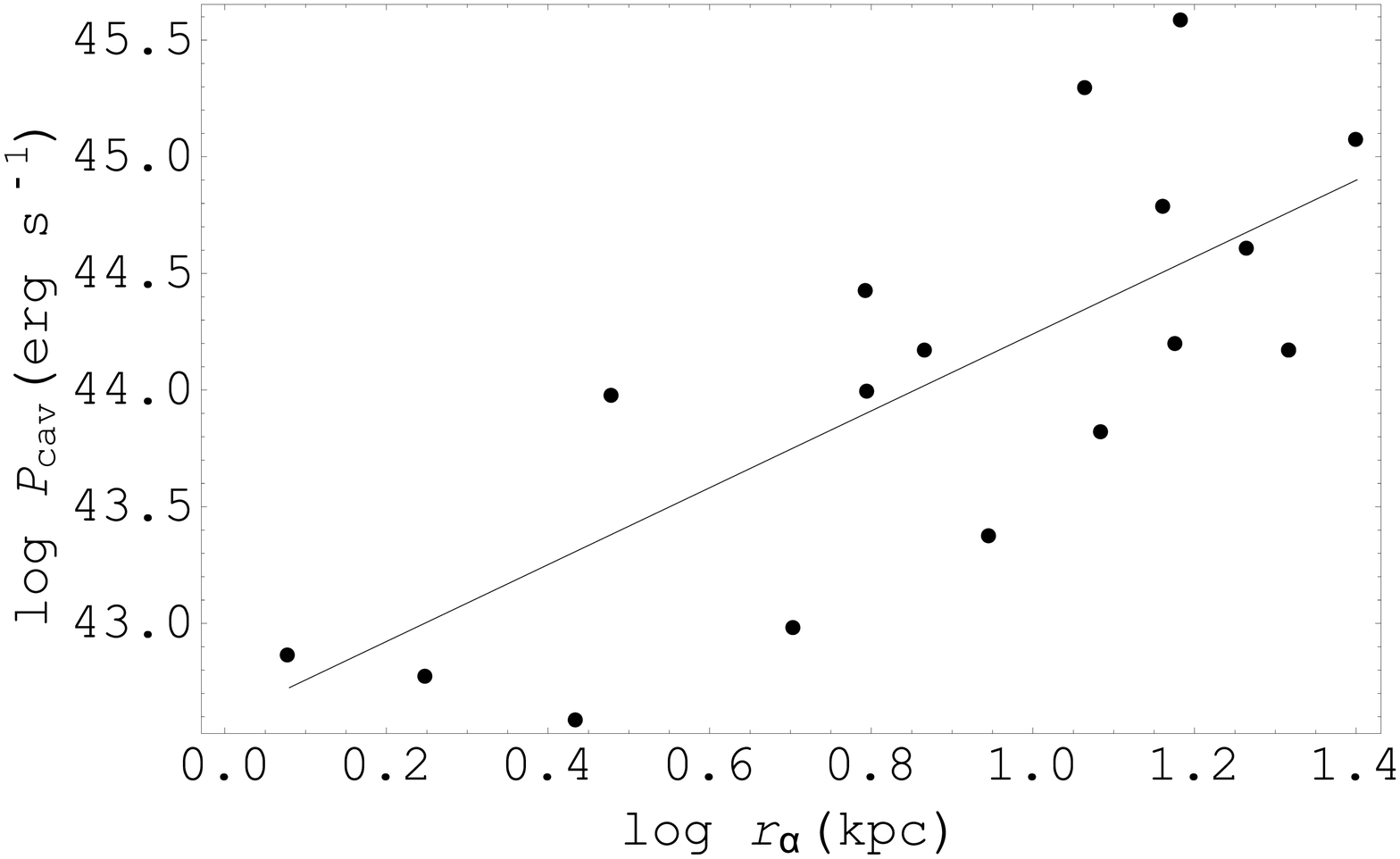}} \\
{\includegraphics[scale=0.28]{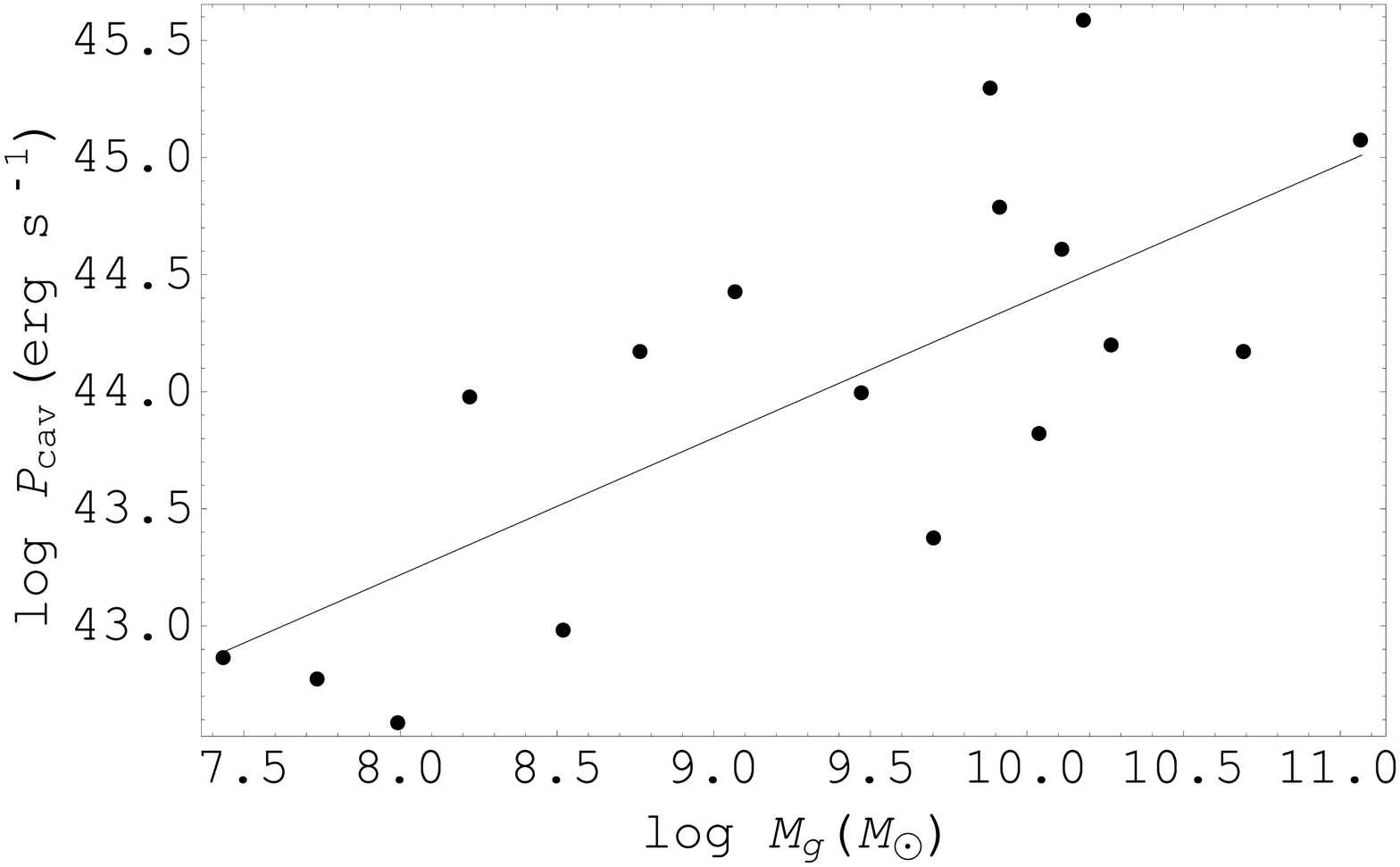}} \\
{\includegraphics[scale=0.28]{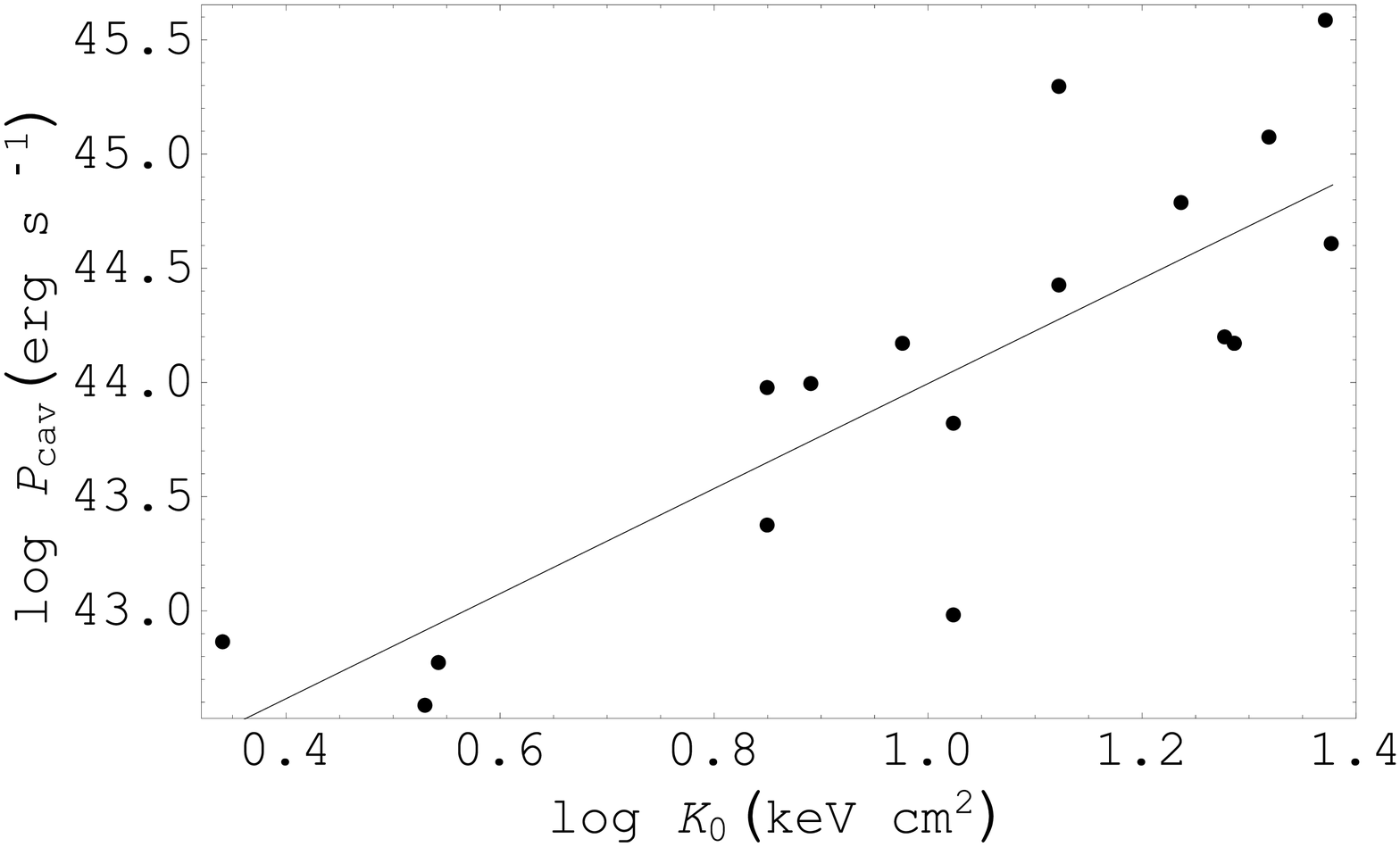}}
\end{tabular}
\caption{The cavity power (see Table 1) versus the size of the flat entropy region
$r_\alpha$ (upper panel), the gas mass parameter $M_g$ as given in equation
(\ref{e-mg}; middle panel), and the entropy floor $K_0$ (lower panel). }
\label{fig:correlations}
\end{figure}
%
%

Two comments are in place here. First, $M_g$ strongly depends on $r_\alpha$, $M_g \propto r_\alpha^3$, hence the
two correlations are not independent of each other. Second, because of the large scatter
and small number of objects, we did not perform any deeper statistical analysis.
We only point to the existence of a correlation, and the potential of
using it to estimate an average, over time and many clusters, SMBH mass accretion rate.
In a forthcoming more detailed study we will try to incorporate elliptical galaxies to the correltion.

With the aid of equation (\ref{e-jets}),
we cast the desired phenomenological formulae for the accretion rate on the SMBH in the form
\begin{equation}
\label{e-corr-d}
\dot M_{\rm acc} \simeq 0.03
\left( \frac{r_\alpha}{10 \kpc} \right)^{1.7}
\left(\frac{\eta_j}{0.1} \right)^{-1}   M_\odot \yr^{-1},
\end{equation}
and
\begin{equation}
\label{e-cormg-d}
\dot M_{\rm acc} \simeq 0.04
\left( \frac{M_g}{10^{10} M_\odot} \right)^{0.6}
\left(\frac{\eta_j}{0.1} \right)^{-1}   M_\odot \yr^{-1}.
\end{equation}
Considering the large uncertainties in the physical parameters and the large
scatter in the graphs, there is no point to give the numerical values in the above expressions
to a higher accuracy.

The accretion rate does not increase as fast as the gas mass parameter $M_g$,
because as we move to larger distances from the center the cooling time increases,
and the gas supply becomes less efficient.
However, equations (\ref{e-corr-d}) and (\ref{e-cormg-d}) support the cold feedback mechanism in
suggesting that mass accreted onto the central SMBH is drained from an extended region,
particularly from where the entropy profile is flat.

\citet{McNa10} find no correlation between the molecular mass in clusters and the cavity power.
We analyzed the 10 clusters that are in our sample and have molecular mass in \citet{McNa10}
(taken from \citealp{edge01}; \citealp{sal03}; \citealp{sal04}; \citealp{sal06}; \citealp{tan08}).
We confirmed their finding that the cavity power has no correlation with the molecular mass
(we get a Pearson R coefficient of $R^2 =0.13$).
We do find a convincing correlation between the molecular mass $M_{\rm mol}$ and the entropy
profile quantities $r_{\alpha}$ and $M_g$, as presented in Fig. \ref{fig:molecular}.
We again look for a simple linear relation in the log-log plots, and find
\begin{equation}
\label{e-molr}
\log M_{\rm mol} (M_\odot) = 2.4 \log r_\alpha({\rm kpc}) + 6.8,
\end{equation}
with Pearson R coefficient of $R^2=0.65$,
and
\begin{equation}
\label{e-molm}
\log M_{\rm mol} (M_\odot) = 0.89 \log M_g (M_\odot) + 0.50,
\end{equation}
with $R^2 =0.71$.
\begin{figure}
\begin{tabular}{c}
{\includegraphics[scale=0.28]{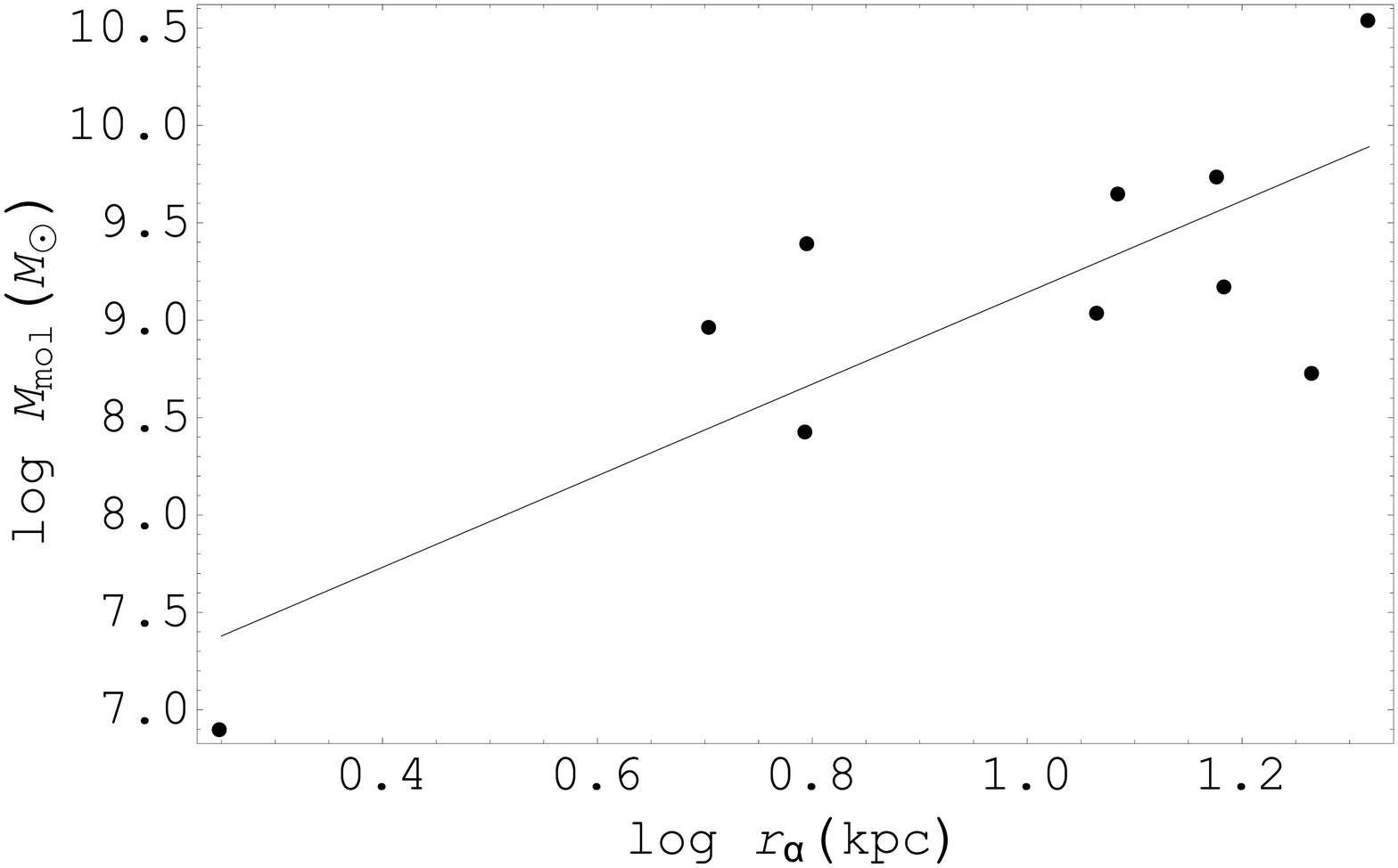}} \\
{\includegraphics[scale=0.28]{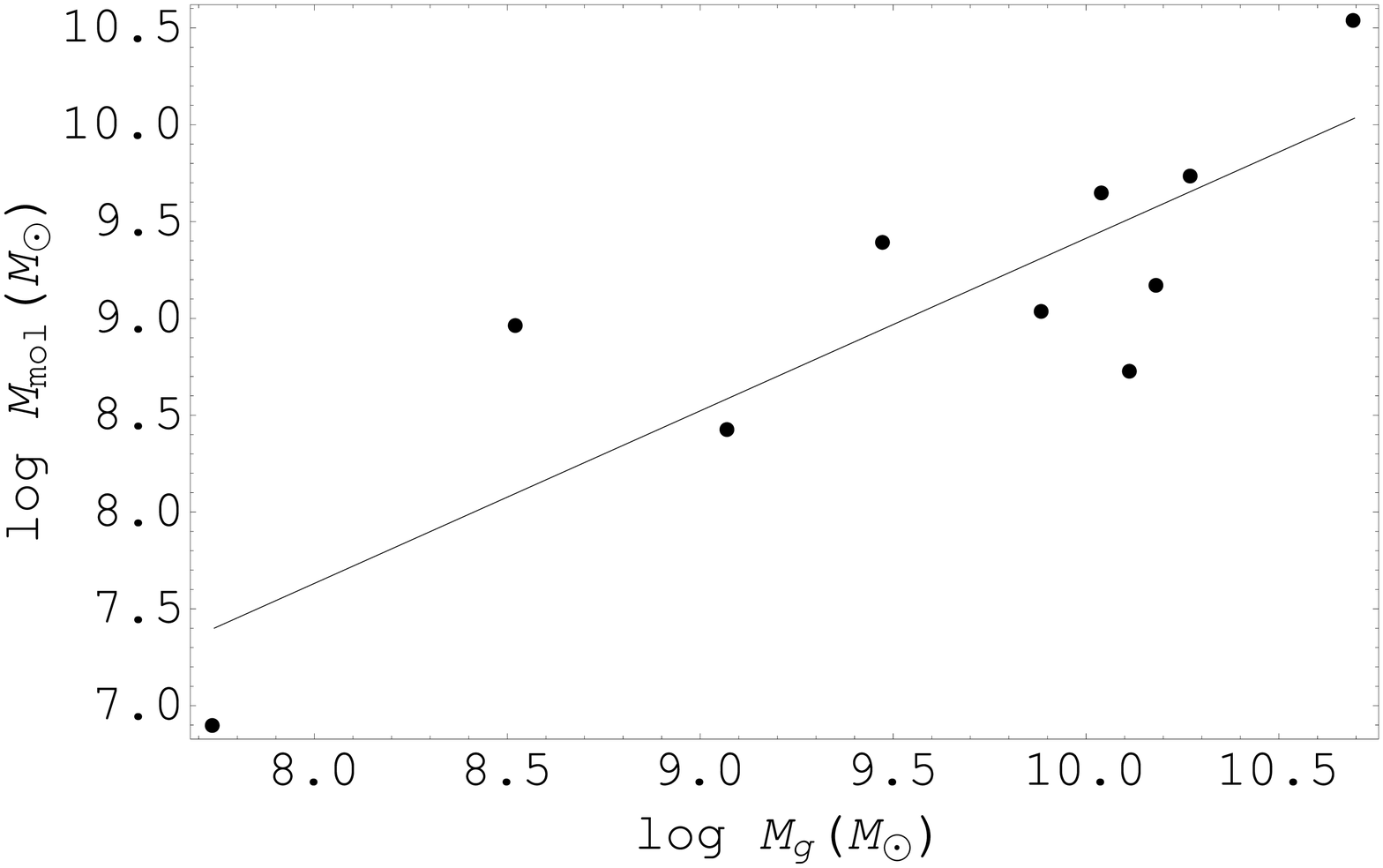}} \\
{\includegraphics[scale=0.28]{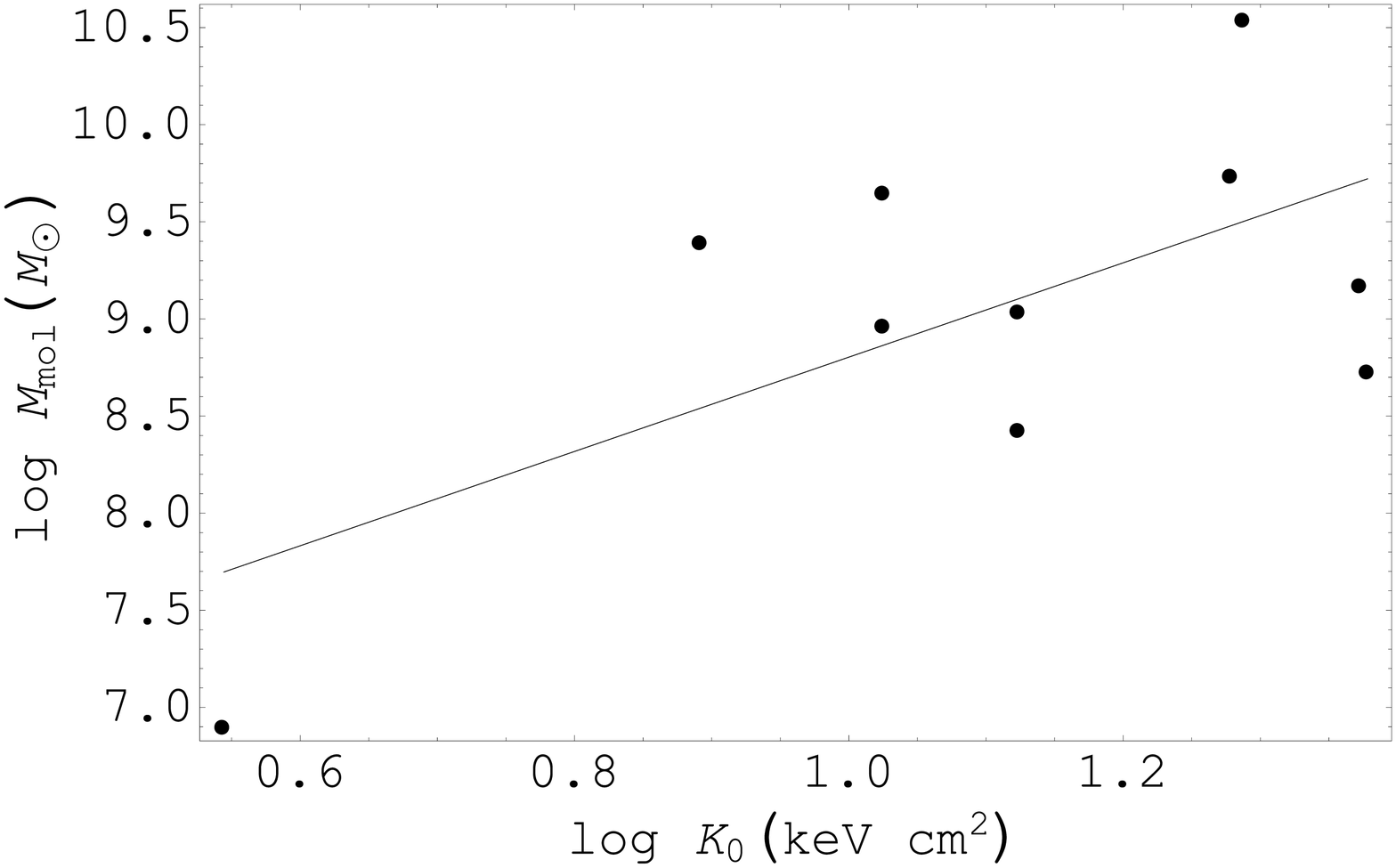}}
\end{tabular}
\caption{The molecular mass versus $r_\alpha$ (upper panel) and versus $M_g$ (middle panel).
For sources of data see Table 1.
}
\label{fig:molecular}
\end{figure}

The above finding is very interesting.
We find correlations of the entropy profile properties with both the cavity power and molecular mass.
However, there is no correlation between the molecular mass and the cavity power.
This shows two things. (1) The situation is not simple. There is a large scatter probably
because the AGN activity substantially varies with time \citep{Nip05}. (2) The entropy
profile seems to be the basic quantity determining the cooling of gas,
both to form cold reservoir and to feed the SMBH.

We also correlate $P_{\rm cav}$ and $M_{\rm mol}$ with the entropy floor $K_0$ (from \citealp{Cav09}),
which is the value of the entropy in the inner flat region.
The correlation with the molecular mass is poor (lower panel of Fig. \ref{fig:molecular}).
As shown in the lower panels of Fig. \ref{fig:correlations},
the jets' (cavity) power clearly increases with the entropy floor.
The linear fitting gives
\begin{equation}
\label{e-corrk0}
\log P_{\rm cav}({\rm erg} \s^{-1}) = 2.3 \log K_0 ({\rm keV} \cm^{2}) + 41.7,
\end{equation}
with $R^2=0.67$.
This behavior is \emph{opposite} to what one would expect from a naive cooling interpretation.
A naive expectation is that for a lower entropy the cooling is more efficient.
However, a higher entropy floor goes along with a more extended (larger $r_\alpha$) region;
we find an almost proportionality relation of $K_0({\rm keV} \cm^{2}) \simeq 1.1 r_\alpha({\rm kpc})$.
We find that the cavity power (and possibly molecular mass) increases with the size of the
flat entropy region. This is expected in the cold feedback mechanism (section \ref{s-basic}),
as long as the cooling time and the entropy floor are below
their respective thresholds of $\tau_{\rm cool} (12 \kpc) < 8.5 \times 10^8 \yr $ \citep{Raf08}
and  $K_0 = kTn^{-2/3} \la 30 \keV cm^{2}$ \citep{Voit08}, respectively.

\section{SUMMARY}
\label{s-summary}

Our main finding is the correlation of the jet power that went to inflate the bubble
$P_{\rm cav}$, with the size of the flat entropy region $r_{\alpha}$ (Fig. \ref{fig:correlations}).
(By flat profile we refer to a moderate profile with $\alpha<0.5$. )
The positive correlation between $P_{\rm cav}$ and properties of the flat entropy
region in the inner regions of clusters is expected in the cold feedback mechanism.
However, at this point the cold feedback mechanism does not give the form of the correlation.
We have tried the simplest correlation in the log-log plot, and derived equations
(\ref{e-corr}), (\ref{e-corr-d}).
The correlation with the gas mass parameter defined in equation (\ref{e-mg}) is given in
equations (\ref{e-cormg}), (\ref{e-cormg-d}).

 There is a slight possibility that the correlation arises from an underlying effect that is
not related to the feedback process. For example, a larger flat entropy region will cause jets
to deposit more of their energy in the inner region and inflate large bubbles.
However, the positive correlation of $r_\alpha$ with the molecular gas (Fig. \ref{fig:molecular}),
despite that the molecular gas does not correlate with the jet power, bring us to reject
this possibility. Further support to this rejection is the positive correlation with the
value of the entropy floor (the flat part at the center), as discussed in the last paragraph
of section \ref{s-SMBH}.
We rather attribute a fundamental role to the flat entropy profile in the inner region
in determining the cooling of the hot gas: The larger the flat entropy region is, the larger is
the clumps' draining volume.

The relations (\ref{e-corr-d}) and (\ref{e-cormg-d}) derived here, and a similar one that can be
derived from equation (\ref{e-corrk0}), can be used for an estimate of the average mass
accretion rate onto the SMBH. They should replace the Bondi accretion that does not fit
accretion onto the SMBH in the center of clusters of galaxies (\citealp{Sok06, Sok09, Piz10,McNa10}).


\end{document}